
\tolerance=10000
\input phyzzx

\font\mybb=msbm10 at 12pt
\def\bbbb#1{\hbox{\mybb#1}}
\def\Z {\bbbb{Z}}
\def\R {\bbbb{R}}




\def \ffi {\phi}

\def \mm {\mu}

\def \ss {\sigma}
\def \tt {\tau}

\def \2 {{1 \over 2}}
\def \3 {{1 \over 3}}
\def \4 {{1 \over 4}}
\def \5 {{1 \over 5}}
\def \6 {{1 \over 6}}
\def \7 {{1 \over 7}}
\def \8 {{1 \over 8}}
\def \9 {{1 \over 9}}
\def \0 { \infty}

\def\++ {{(+)}}
\def \- {{(-)}}
\def\+-{{(\pm)}}

\def\ek {\eqn\abc$$}

\def \qq {\qquad}


 \def\unit{\hbox to 3.3pt{\hskip1.3pt \vrule height 7pt width .4pt \hskip.7pt
\vrule height 7.85pt width .4pt \kern-2.4pt
\hrulefill \kern-3pt
\raise 4pt\hbox{\char'40}}}

\def\cM {{\cal{M}}}

\def\nup#1({Nucl.\ Phys.\  {\bf B#1}\ (}


\REF\Rom{ L.J.~Romans,
                 {  Phys.~Lett.}~{\bf 169B} (1986) 374.}
\REF \ShSh{J. Scherk and J.H. Schwarz,     {  Nucl. Phys.}~{\bf B153} (1979)
61.}
\REF\noel{ H. Nicolai, P.K. Townsend and P. van Nieuwenhuizen,   Nuovo
Cim.Lett. {\bf 30} (1981) 315;
K.~Bautier, S.~Deser, M.~Henneaux and
                  D.~Seminara,
                  {  Phys.~Lett.}~{\bf 406} (1997) 49,
                 {  hep-th/9704131};
                  S.~Deser,
                   {  hep-th/9712064},
                  {  hep-th/9805205}.}
\REF\bergy{E. Bergshoeff, M. de Roo, M.B. Green, G. Papadopoulos and P.K.
Townsend, Nucl. phys. {\bf B470} (1996) 113.}
\REF\west{P. S. Howe,  N. D. Lambert,  P. C. West, Phys.Lett. B416 (1998) 303;
hep-th/9707139.}
\REF\fibs{ I.V.~Lavrinenko, H.~Lu and C.N.~Pope,   hep-th/9710243.}
\REF\berlozgort{E. Bergshoeff, Y. Lozano and T. Ortin,
Nucl.Phys. B518 (1998) 363-423,
hep-th/9712115.}
\REF\Pol{J. Polchinski, Phys. Rev. Lett. 75 (1995) 4724; hep-th/9510017 .}
\REF\asp{P. Aspinwall, Nucl. Phys. Proc. Suppl. {\bf  46}
  (1996) 30, hep-th/9508154; J.H. Schwarz, hep-th/9508143.}
\REF\ScherkSchwarzother{
E. Cremmer, J. Scherk and J.H. Schwarz, Phys.Lett. {\bf 84B} (1979) 83;
see also T.R. Taylor in SUSY 95 ed. I. Antoniadis and H. Videau, Frontieres
(1986) 389. }
\REF\ScherkSchwarzothers{
S. Thomas and P.C. West, Nucl. Phys. {\bf 245} (1984) 45;
M. Porrati and F. Zwirner, Nucl. Phys. {\bf B326} (1989) 162.}
\REF\ScherkSchwarzothert{R. Rohm, Nucl. Phys. {\bf B237} (1984) 553;
C. Kounnas and M. Porrati, Nucl. Phys. {\bf B310} (1988) 355;
S. Ferrara, C. Kounnas, M. Porrati and F. Zwirner, Nucl. Phys.
{\bf B318} (1989) 75;
S. Ferrara, C. Kounnas and M. Porrati, Phys. Lett. {\bf B206}
(1988) 25.}
\REF\wall{ P.M. Cowdall,  H. Lu,  C.N. Pope,  K.S. Stelle,
P.K. Townsend,
Nucl.Phys. B486 (1997) 49; hep-th/9608173 .}
\REF\gage{ E. Bergshoeff ,  M. de Roo ,  E. Eyras, Phys.Lett. B413 (1997)
70-78; hep-th/9707130.}
\REF\kuri{ N.~Kaloper, R.R.~Khuri and R.C.~Myers,
                {  hep-th/9803006}.}
\REF\fibs{ I.V.~Lavrinenko, H.~Lu and C.N.~Pope,   hep-th/9710243.}
\REF\ort{ P.~Meessen and T.~Ortin,  {  hep-th/9806120}.}
\REF\CJ{E. Cremmer and B. Julia, Phys. Lett. {\bf 80B} (1978) 48; Nucl.
Phys. {\bf B159} (1979) 141.}
\REF\julia{B. Julia in {\it Supergravity and Superspace}, S.W. Hawking and M.
Ro$\check c$ek, C.U.P.
Cambridge,  (1981). }
\REF\HT{C.M. Hull and P.K. Townsend, hep-th/9410167.}
\REF\fvaf{C. Vafa, Nucl. Phys. {\bf 469} (1996) 403.}
\REF\Fm{
D. Morrison and C. Vafa, Nucl. Phys. {\bf B473} (1996) 74,
hep-th/9602114; Nucl. Phys. {\bf B476} (1996) 437, hep-th/9603161;
 A. Sen, Nucl. Phys. {\bf B475} (1996) 562, hep-th/9605150;
R. Friedman, J. Morgan and E. Witten, Comm. Math. Phys. {\bf 187} (1997)
679, hep-th/9701162; M. Bershadsky, A. Johansen, T. Pantev and V. Sadov, Nucl.
Phys.
{\bf B505} (1997) 165, hep-th/9701165.}
\REF\Sen{A. Sen, hep-th/9802051.}
\REF\genvaf{A. Kumar  and C. Vafa,  Phys.Lett.  {\bf  B396}  (1997) 85;
hep-th/9611007.}
\REF\vafwit{C. Vafa and E. Witten, Nucl.Phys.Proc.Suppl. 46 (1996) 225;
hep-th/9507050.}
\REF\bergort{E. Bergshoeff, C.M. Hull and T. Ortin,
 Nucl. Phys. {\bf B451} (1995) 547,hep-th/9504081.}
\REF\berr{E. Bergshoeff,  M. de Roo,  E. Eyras,  B. Janssen,  J. P. van der
Schaar
Class. Quant. Grav. {\bf 14} (1997) 2757;
hep-th/9704120.}
\REF\GravDu{C.M. Hull, Nucl. Phys. {\bf B509} (1997) 252, hep-th/9705162.}
\REF\bergnin{ E.~Bergshoeff and J.P.~van der Schaar,
                {  hep-th/9806069}.}
\REF\hulmat{C.M. Hull,  hep-th/9712075.}




\Pubnum{ \vbox{  \hbox {QMW-PH-98-36} \hbox{LPTENS 98/32}
\hbox{hep-th/9811021}} }
\pubtype{}
\date{November, 1998}

\titlepage

\title {\bf  Massive String Theories From M-Theory and F-Theory}

\author{C.M. Hull}
\address{Physics Department, Queen Mary and Westfield College,
\break Mile End Road, London E1 4NS, U.K.}

\abstract {The massive IIA string theory whose low energy limit is the massive
supergravity theory
constructed by Romans is obtained from M-theory compactified on a 2-torus
bundle
over a circle in a
limit in which the volume of the bundle shrinks to zero. The massive string
theories in
9-dimensions
given by Scherk-Schwarz reduction of IIB string theory
are  interpreted as F-theory compactified on
2-torus bundles over a circle.  The M-theory solution that gives rise to the
D8-brane  of the massive IIA theory is identified. Generalisations of
Scherk-Schwarz reduction
are discussed.
 }

\endpage

There is a massive version of the ten dimensional type IIA supergravity due
to Romans
[\Rom] and it has long been a mystery as to whether it has an eleven
dimensional
origin, in which the mass might arise from an explicit mass in eleven
dimensions,
or from a parameter in the dimensional reduction ansatz, as in
Scherk-Schwarz  dimensional
reduction  in which the fields have non-trivial dependence on the coordinates
of the  internal dimensions
[\ShSh].
 In [\noel], it has been argued, subject to certain assumptions, that no
covariant
massive deformation of 11-dimensional supergravity is possible, which would
mean that  the massive IIA supergravity cannot come from a conventional
reduction of
such a massive theory.
Dimensionally reducing the Romans
theory on a circle gives a massive 9 dimensional theory which can also be
obtained from the type IIB theory by a Scherk-Schwarz
reduction   [\bergy], but
the Romans theory cannot be obtained by a  Scherk-Schwarz reduction of
11-dimensional supergravity; a different 10-dimensional massive supergravity
theory, for which there
is no action, was proposed in [\west], and obtained via
a  Scherk-Schwarz reduction of 11-dimensional supergravity
in [\fibs].
In [\berlozgort], the Romans supergravity was lifted to a massive
deformation of 11-dimensional supergravity where the terms in the
11-dimensional
action
 depending on the mass parameter $m$ also depend explicitly on the Killing
vector used in the
dimensional reduction to 10-dimensions, and so this 11-dimensional theory is
not fully covariant.

The IIA supergravity is the field theory limit of the IIA superstring, and the
strong coupling limit of the IIA superstring is M-theory, which  has
11-dimensional supergravity as its field theory limit.
There is a
massive version of type IIA string theory [\Pol]
whose
field theory
limit is the Romans supergravity theory (see for instance [\bergy]),
and the question arises as to how this
massive IIA string theory arises from M-theory.
 Our purpose here is to argue that although the Romans
supergravity theory may not be derivable from
 11-dimensional supergravity, or any covariant
massive deformation
thereof, the massive IIA superstring, whose low energy limit is the Romans
theory, can be obtained
from M-theory.

The type IIB supergravity  theory also cannot be obtained from 11-dimensional
supergravity, but
the type IIB string  theory can  be obtained from M-theory by compactifying on
a 2-torus and taking a
limit  in which
the area of the
torus tends to zero while the modulus $\tau$ tends to a constant, the imaginary
part of which is the string  coupling constant of the IIB string theory
  [\asp].
The massive IIA string theory compactified on a circle of radius $R$
is T-dual to a Scherk-Schwarz compactification of the IIB superstring on a
circle of radius $1/R$,
with mass-dependent modifications of  the usual T-duality rules
[\bergy].
Thus the massive IIA string can be obtained from M-theory by first reducing on
a 2-torus that shrinks to zero size to obtain the IIB string, and then
using  a \lq twisted' T-duality to obtain the massive IIA string, by making a
Scherk-Schwarz  reduction on a circle and then shrinking the radius to zero
size.
Moreover, we shall argue that the Scherk-Schwarz compactification of the IIB
superstring
 has a natural formulation in terms of F-theory.  The Scherk-Schwarz reduction
of the IIB string theory can  then be obtained from a limit
of a compactification of
M-theory, using the relation between F-theory and M-theory, and it will be
shown that the massive IIA string can be obtained by reducing M-theory on a
torus bundle over a circle and taking a limit in which the bundle shrinks to
zero size, with all three radii tending to zero.
It will be seen  that this relates the D8-brane, which
only occurs in the IIA string with non-vanishing mass, to
a brane-like solution
of M-theory, which might be thought of as an M9-brane, and to a related
12-dimensional F-theory \lq solution'.

The Scherk-Schwarz mechanism  and its generalisations
[\ShSh,\ScherkSchwarzother-\ort] introduces
mass parameters   into toroidal  compactifications of
supergravities and string
theories. If the original theory has a global symmetry $G$ acting on fields
$ \phi$ by $ \phi\to g( \phi)$, then in a  generalised Scherk-Schwarz
reduction or twisted reduction the
fields are not independent
of the internal coordinates, but are chosen to
depend on the torus coordinates $y$ through an ansatz
$$ \phi (x^\mm,y) = g_y(\ffi(x^\mm))
\eqn\ansa$$
 for some $y$-dependent symmetry transformation $g_y=g(y)$ in $ G$.
In many cases  this leads to a spontaneous breaking of supersymmetry [\ShSh],
while in others  it results in the gauging of certain symmetries of   the
conventionally reduced theory, and the introduction of
a scalar potential and cosmological constant [\wall,\gage,\ort].
Here, we will restrict ourselves to compactifications on a circle, with
periodic coordinate $y \sim y+1$.
For example, for reducing a theory with a linearly realised $U(1)$ symmetry on
a circle, a massless
field $\phi$ of charge
$q$ can be given a $y$ dependence
$\phi(x,y)= e^{2\pi iqmy}\ffi(x)$, so that  the field $\ffi(x)$ is given  a
mass of $qm$.

The map $g(y)$ is not periodic, but
 has
 a {\it monodromy}
$$
{\cal M} (g)= g(1)g(0)^{-1}
\eqn\abc$$
for some ${\cal M}$ in $G$. We will consider here maps of the form
$$g(y)= \exp (My)
\eqn\ansatz$$
for some
 Lie algebra element $M$, so that
the monodromy is
$$
{\cal M} (g)= \exp M
\eqn\mono$$
Then
$$ M=g^{-1}\partial _y g
\ek
is proportional to the mass matrix of the dimensionally reduced theory
and is independent of $y$ [\ort].

The next question is whether two different choices of $g(y)$ give inequivalent
theories.
The   ansatz breaks the symmetry $G$ down to the subgroup
preserving
$g(y)$, consisting of those $h$ in $G$ such that $h^{-1} g(y) h=g(y)$.
Acting with a general constant element  $k$ in $G$ will change the
mass-dependent
terms, but will give a
$D-1$ dimensional theory related to the original one via the field refinition
$\ffi \to k(\ffi)$. This same theory could have been obtained directly via a
 reduction using
$k^{-1} g(y) k$ instead of $g(y)$, so two choices of $g(y)$ in the same
conjugacy class
give equivalent reductions (related by field-redefinitions). As a result, the
reductions are
classified by  conjugacy classes of the
mass-matrix $M$.

The map $g(y)$ is a  local
section of a principal fiber  bundle over the circle with fibre $G$
and monodromy
${\cal M}  (g)$ in $G$.
Such a bundle is constructed from $I\times G$, where $I=[0,1]$ is the unit
interval, by gluing the
ends of the interval together with a twist of the fibres by the monodromy
${\cal M} $.
Two such bundles with monodromy in the same $G$-conjugacy
class are equivalent.
Only those monodromies  $\cM$ that can be written as $e^M$ for some $M$ arise
in this way, and for
those monodromies that are in the image of the exponential map,   there are in
general an infinite
number of  possible choices of mass-matrix. Indeed, if
 $M,M'$ are two such mass matricies for a given monodromy such that
$e^M=e^{M'}=\cM$,
then $e^M e^{-M'}= 1$ and so there is a $\lambda$ satisfying $e^\lambda=1$ such
that
$M-M'-{1 \over 2}[M,M']+\dots = \lambda$.
The   general solution of $e^M=\cM$ is then of the form
 $M=M'+\lambda+{1 \over 2}[M',\lambda]+...$ where $M'$ is a particular solution
and $\lambda$ is any solution of $e^\lambda=1$. The   algebra elements
$\lambda$ with $e^\lambda=1$
fall into adjoint orbits, as,  for any group element $g$,  $\lambda ' = g
\lambda g^{-1}$ satisfies this condition if $\lambda$ does.
The set of all Lie algebra elements $\lambda$ with $e^\lambda=1$
is given by  the  adjoint orbits  of all points in
 the   dual of the weight lattice of the maximal compact subalgebra $H$ of $G$,
sometimes called the integer lattice.

Of particular interest are the    D-dimensional supergravity theories with
rigid duality symmetry $G$ and scalars
taking values in
$G/H$ [\CJ,\julia], which  can be  Scherk-Schwarz-reduced on a circle to $D-1$
dimensions.
The reduction requires the choice of a map $g(y)$ of the form \ansatz\ from
$S^1$ to $G$, which then determines the $y$-dependence of the fields   through
the ansatz
\ansa, and any choice of Lie algebra element $M$ is allowed.
In the quantum theory, the symmetry group $G$ is broken to a discrete sub-group
$G(\Z)$ [\HT]. A consistent twisted reduction of a string or M-theory,
whose
low-energy
effective theory is the supergravity theory considered above, then
requires that the monodromy be in the U-duality group $G(\Z)$.
(In the classical supergravity
theory, any
element of $G$ can be used as the monodromy.)
Then
  the
choice of $M$ is restricted by the constraint that $e^M$ should be in $G(\Z)$.
As before,
if $M=kM'k^{-1}$ where $k$ is in $G$, the theories are related by field
redefinitions. However,
 only if $k$ is in $G(\Z)$ will the redefinition preserve the charge lattice
[\HT].  Once the conventions for the definitions of charges are fixed, it is
necessary
to restrict to conjugation by elements of
  $G(\Z)$, and so
reductions are specified
by $G(\Z)$ conjugacy classes  of maps \ansatz\ with monodromy \mono\ in
$G(\Z)$.

Here we will concentrate   on the examples relevant to the massive IIA
superstring. The type IIB supergravity
theory has $G=SL(2,\R)$ global symmetry and any element $M$ of the
$SL(2,\R)$  Lie algebra
 can be used
in the ansatz \ansa,\ansatz\ to give a Scherk-Schwarz reduction to 9-dimensions
to
obtain a class of massive
9-dimensional supergravity theories.
Such reductions for particular elements of $SL(2,\R)$ were given in
[\bergy,\kuri,\fibs],  and the   general class of $SL(2,\R)$
reductions of IIB
supergravity was obtained   in  [\ort].
Note that this ansatz does not allow the monodromy to be an arbitrary
$SL(2,\R)$
group element, but requires it to be in the image of the exponential map.
Acting with an $SL(2,\R)$ transformation leaves the mass-independent part of
the theory
unchanged
but changes the mass matrix  by
$SL(2,\R)$
conjugation, and so there are three distinct classes of inequivalent theories,
corresponding to the
hyperbolic, elliptic and parabolic $SL(2,\R)$ conjugacy classes, represented by
monodromy matrices of the form
$$\pmatrix{
a & 0 \cr
0 & a^{-1} \cr
}, \qq
\pmatrix{
\cos \theta & \sin \theta \cr
- \sin \theta & \cos \theta \cr
}
, \qq
\pmatrix{
1 & a \cr
0 & 1 \cr
}
\ek
respectively.
The details of the  reduction of the bosonic sector of the supergravity theory
for general $M$ were given in [\ort].

In the quantum theory, only an $SL(2,\Z) $ symmetry  remains [\HT].
The quantum-consistent Scherk-Schwarz reductions of this theory
to 9 dimensions
are those for which the
monodromy is in $SL(2,\Z)$,
 and are defined up to
$SL(2,\Z)
$ conjugacy.
The fact that the monodromy must be in $SL(2,\Z)$ implies a quantization of the
masses.

The IIB supergravity scalars take values in $SL(2,\R)/U(1)$ and can be
represented by a complex scalar $\tt=C_0+ie^{-\Phi}$  transforming under
$SL(2)$ by fractional linear
transformations, so that
$g \in SL(2)$ acts as $$g=\pmatrix{
a & b \cr
c & d \cr
}: \tau \to  \tau_g = {a\tt +b\over c \tt +d}
\ek
 The
Scherk-Schwarz ansatz,   $\tt (x,y) =\tau (x)_{g(y) }$, gives a
complex scalar $\tau (x)_{g(y) }$ of the reduced theory and
for fixed $x$, $ \tau (x)_{g(y) }$ depends on $y$ and
  is a section of the
bundle over the circle with fibre  $SL(2,\R)/U(1)$ obtained as a quotient of
the principle $SL(2,\R)$ bundle by $U(1)$.

For the IIB string theory, the monodromy must be restricted to lie in
$SL(2,\Z)$.
If $g(y)$ has   $SL(2,\Z)$ monodromy, the local section $\tt (y) =\tau _{g(y)
}$
  can be used to construct a torus bundle over the circle in which $\tt$ is the
$T^2$ modulus, and depends on the position on the circle.
The total space of the torus bundle is  a 3-dimensional space  $B$  with metric
$$ds_B^2=
R^2 dy^2
+{A \over Im (\tau)}\vert dz_1 +\tt (y) dz_2 \vert ^2
\eqn\bunmet$$
where
the fibre is a $T^2$ with real periodic coordinates $z_1,z_2$, $z_i\sim z_i
+1$, constant area modulus $A$ and
complex structure $\tt(y)$, which depends on the coordinate $y$ of the circular
base space, and this
has circumference $R$.
The Scherk-Schwarz reduction of the IIB superstring  with an ansatz $\tt (y)
=\tau _{g(y) }$
associated with a
particular torus
bundle $B$ is precisely
what is meant by F-theory compactified on the three dimensional total space $B$
[\fvaf-\genvaf].

This generalises;
for theories in which the global symmetry is $G=SL(n,\R)$ with quantum symmetry
$SL(n,\Z)$, a twisted  reduction on an $m$-torus in which all monodromies
are in $SL(n,\Z)$
corresponds to a torus bundle with fibres $T^n$ over a base $T^m$. For $m=1$,
this gives a $T^n$ bundle over a circle.
Certain torus bundles over a circle
 are also circle bundles over a torus, and the latter
was the interpretation used in [\fibs]. However, the
torus bundle over a circle is both more general  and more useful, as it has an
F-theory
interpretation.
For example, the 7-dimensional maximal supergravity
theory has
$G=SL(5,\R)$
symmetry, while the 7-dimensional type II string theory has $SL(5,\Z)$
U-duality.
The general twisted reduction from 7  to 6 dimensions would involve a
map
$g(y):~S^1 \to SL(5,\R)$ with $SL(5,\Z)$ monodromy, which is also the data for
a $T^5$ bundle over $S^1$. Then the general $SL(5)$ Scherk-Schwarz reduction
can be
re-interpreted as a
reduction of
the $F'$-theory of  [\genvaf] on a $T^5$ bundle over $S^1$.
(The  $F'$-theory is an analogue of F-theory, also in 12 dimensions,
 which can be compactified on spaces admitting a $T^5$ fibration [\genvaf].)

Any  twisted reduction  of the IIB string to 9 dimensions can be
recast as the
reduction
of F-theory on a bundle $B$ which is a $T^2$ bundle over $S^1$.
One can also consider compactifications of M-theory on $B$, and the two are
related by fibre-wise duality as follows.
For M-theory compactified on $B$ in which the $T^2$ fibres have a constant area
$A$, the limit
$A\to 0$ keeping the modulus
$\tau(x,y)$  fixed
gives F-theory compactified on $B$ with fixed torus area $A=1$, say.
For a trivial bundle, this follows from the fact that M-theory compactified on
 $T^2$ becomes, in
the limit in which the torus shrinks to zero size, the IIB string theory, and
the generalisation to
non-trivial bundles follows
from the adiabatic argument [\vafwit].

Consider the  Scherk-Schwarz reduction using the map $S^1\to SL(2,\R)$
$$ g(y)=
\pmatrix {
1 &
my \cr
0 & 1}, \qq  M=
\pmatrix {
0 &
m \cr
0 & 0}
 \eqn\abc$$
so that    \ansa\ leads to the linear ansatz
$$
\tt(x,y)=\tt(x) + my
\eqn\ttis$$
The monodromy is
$$ {\cal M}  =
\pmatrix {
1 &
m \cr
0 & 1}
 \eqn\monod$$
and in the quantum theory this must be in $SL(2,\Z)$  so that   $m$ must be an
integer,  and the
mass is quantized, as it is proportional to $m$.
This is precisely the reduction studied in [\bergy], and is T-dual to the
massive IIA string theory,
with mass parameter $m$, conventionally compactified on $S^1$.
The bundle $B$ has a metric given by \bunmet,\ttis, which takes the simple form
$$
ds^2=   dy^2 + (dz_1 +mydz_2)^2 + dz_2^2
\eqn\abc$$
 if $\tt_0=i, A=R=1$. This 3-space $B$ is also a circle bundle over a 2-torus
with
fibre coordinate $z_1$, base-space coordinates $y,z_2$ and connection 1-form
${\bf A}=mydz_2$ [\fibs].

The massive IIA string theory arises from M-theory as follows.
Let $B(A,R)$ be the
the torus bundle   over a circle of radius $R$, where the torus has
modulus
$\tt$ depending on the $S^1$ coordinate $y$ through
$$\tt= \tt_0+my
\eqn\yttr$$
for some constant $\tt_0$, and $y$-independent area $A$.
Compactifying M-theory on $B(A,R)$ and taking the limit $A\to 0$ gives F-theory
compactified on
$B(1,R)$, or equivalently the Scherk-Schwarz reduction of the IIB string on a
circle of
radius $R$ using the ansatz \ttis.
This is T-dual to the massive IIA string with mass parameter $m$ compactified
on a circle of radius
$1/R$, and so the uncompactified massive IIA string is obtained by taking the
limit $R\to 0$.
Putting this together, we obtain the massive IIA string by compactifying
 M-theory on $B(A,R)$ and taking the zero-volume limit $A\to 0$, $R\to 0$.
The bundle also depends on $\tt_0$ and $m$, and is trivial if $m=0$, in which
case
$B$ is a 3-torus, and M-theory on a 3-torus indeed gives, in the limit in which
the torus shrinks
to zero size, the massless IIA string theory or M-theory, depending on the
value of the
string coupling. The IIB string coupling constant $g_B$ is given by the
imaginary
part of $\tt_0$, $g_B=
1/Im(\tt_0)$, and the coupling constant $g_A$ for the T-dual IIA theory
is related to this by $g_A=g_B /R$, so that
$$g_A= {1\over Im(\tau_0)R}
\eqn\tertte$$
Then if $Im(\tau_0) \to \infty $ as $R \to 0$ so that $Im(\tau_0)R$ remains
fixed, the massive IIA theory at finite string coupling \tertte\ is obtained.
The massive IIA string theory can also be obtained from F-theory on $B(1,R)$ by
taking the limit $R
\to 0$, keeping $Im(\tau)R$ fixed.

The massive IIA supergravity theory doesn't have a Minkowski or (anti) de
Sitter solution, and
there is no maximally supersymmetric solution.
There is a D8-brane solution which preserves half of the supersymmetries,
however [\bergy]. The
string-frame metric is
$$
ds^2=H^{-1/2} d\ss_{8,1} ^2 +H^{1/2}dx^2
\eqn\eig$$
where $d\ss_{p,1} ^2$ is the $p+1$ dimensional Minkowski metric on $\R^{p,1}$.
There is an 8+1 dimensional longitudinal space and a one-dimensional transverse
space with
coordinate
$x$. The function $H(x)$ is harmonic, $H''=0$, and the solution
$$H=\cases{c+m'|x|
 & for $x<0$ \cr
c+m|x| & for $x>0$
}
\eqn\sevdfdsf$$
for some constant $c$ represents a
domain wall at $x=0$, separating regions with two different (integer) values of
the mass parameter, $m$ and $m'$.
If one of the longitudinal coordinates, $y$ say, is made periodic,   a
T-duality in the
$y$-direction leads to the circularly symmetric IIB D7-brane solution of
[\bergy], with string-frame
metric
$$
ds^2=H^{-1/2} d\ss_{7,1} ^2 +H^{1/2}(dx^2+dy^2)
\eqn\sev$$
and
$$e^{-\phi}=H,\qq C_0'=H'
\eqn\abc$$
where $\ffi$ is the dilaton and $C_0$ is the RR scalar.
In Einstein frame, the metric is
$$
ds^2= d\ss_{7,1} ^2 +H (dx^2+dy^2)
\eqn\seve$$
Dimensional reduction in the $y$ direction of the D8-brane \eig\ or D7-brane
\sev\ leads to the
7-brane solution [\wall] of the massive 9-dimensional theory (obtained by
twisted reduction
of the IIB theory
using \ttis) with metric
$$
ds^2=H^{-1/2} d\ss_{7,1} ^2 +H^{1/2}dx^2
\eqn\sevn$$

Conventional dimensional reduction of 11-dimensional supergravity on a 2-torus
gives
massless
9-dimensional type II theory with scalars in the coset space
$\R^+\times SL(2,\R)/U(1)$, which is the moduli space of the torus [\bergort].
A Scherk-Schwarz reduction of this to 8-dimensions
using the ansatz  \ttis\ for the complex   scalar  in $SL(2,\R)/U(1)$
gives a
  massive type II supergravity in 8-dimensions [\wall] and this theory has a
6-brane solution [\wall]
with metric
$$
ds^2=H^{2/3}\left (
H^{-1/2} d\ss_{6,1} ^2 +H^{1/2}dx^2
\right)
\eqn\six$$
However, this massive 8-dimensional theory  arises directly from reduction from
11-dimensions
on the torus bundle $B$,
and we will now check that the  6-brane solution arises from an 11-dimensional
solution reduced on
the torus bundle $B$.
The moduli $\tt,A,R$ of the bundle become scalar fields in the
dimensionally reduced theory, and
for the 11-dimensional oxidation of the solution \six, these moduli
 can be expected to be functions of transverse coordinate $x$.
The 11-dimensional oxidation of \six\ was   given in [\wall,\fibs,\berr], with
metric
$$
ds^2=  d\ss_{6,1} ^2 +Hdx^2 +H(dy^2 + Adz_2^2) +AH^{-1}(dz_1+mydz_2)^2
\eqn\mnin$$
where $A$ is a constant that can be absorbed into a rescaling of $z_1,z_2$.
This can   be rewritten in the form
$$
ds^2=H^{1/2}\left (
H^{-1/2} d\ss_{6,1} ^2 +H^{1/2}dx^2
\right) + ds_B ^2
\eqn\chas$$
where
$  ds_B^2$ is a $B$-metric of the form \bunmet,\ttis, but where the moduli
$\tau,R$
depend on $x$ as well as $y$:
$$
R=H^{1/2} ,   \qq
\tt=my+iH,
\eqn\mods$$
The metric is of the form
$\R^{6,1}\times M_4$
where $M_4$ is of the form $\R \times B$ with coordinates $x,y,z_1,z_2$ and
Ricci-flat metric
$$
ds^2= Hdx^2
  + ds_B ^2
\eqn\abc$$
with the moduli of $B$   given by \mods.
The 11+1 dimensional space $\R^{7,1}\times M_4$
is Ricci-flat and is the F-theory \lq solution' that gives rise to the
Einstein-frame 7-brane
solution  \seve, which can be reduced
further to the 9-dimensional 7-brane \sevn.
Note that for domain walls separating regions of mass $m,m'$, as in \sevdfdsf,
then there are two different bundles $B,B'$ arising on either side of the wall,
one with monodromy \monod\ and one with monodromy given by \monod\ with $m$
replaced by $m'$.

Now taking the limit in which the total spaces $B,B'$ shrink  to zero size, the
solution  \chas\
becomes the D8-brane solution of the massive IIA string, while taking the limit
in which the $T^2$ fibres shrink to zero size ($A \to 0$) gives the circularly
symmetric D7-brane \sev. This can be seen in a number of ways.   For example,
first dimensionally reducing in the $z_1$ direction
and Weyl rescaling to obtain the IIA string-frame metric,
 \mnin\
becomes the D6-brane solution
$$
ds^2=  H^{-1/2}d\ss_{6,1} ^2 +H^{1/2}(dx^2 +dy^2 + Adz_2^2)  \eqn\mnina$$
where the harmonic function depends only on $x$, so that this can be thought of
as a D6-brane \lq smeared' over the $y$ and $z_2$ directions.
Thus regarding $B$ as a circle bundle over $T^2$ with fibre coordinate  $z_1$,
we can shrink
the fibre to obtain the smeared D6-brane solution of the  IIA theory with
charge proportional to $m$.
Now  the limit $A \to 0$ is obtained by  T-dualising in the $z_2$ direction,
using the rules of [\bergort],
gives the circularly symmetric D7-brane \sev\ of the IIB theory. A further
T-duality in
 the $y$ direction gives the D8-brane solution \eig.
Then taking the limit of  \mnin\
 in which the $T^2$ fibres are shrunk is given by
first reducing on $z_1$ to obtain \mnina\ and then T-dualising in the $z_2$
direction to obtain the D7-brane \sev, while the limit in which the total space
shrinks is given by making a further T-duality in the $y$ direction to obtain
the D8-brane \eig.

In [\GravDu], it was argued that there should be an \lq M9-brane'
that gives rise to the D8-brane of the IIA theory,
arising as a domain wall
in M-theory, and in [\bergnin,\hulmat], such branes were considered
further. In particular,
 in [\hulmat]
it was shown that such branes could not be $SO(9,1)$ invariant, but that   one
of the directions was
special, in the same way that the KK monopole solution giving rise to the
D6-brane is not $SO(7,1)$
invariant, and has a special compact direction corresponding to the Taub-NUT
fibre.
  The solution \mnin\ is a domain wall solution of M-theory that gives the
D8-brane of the
massive IIA theory in the limit in which the 3-space B shrinks to zero size,
and so might be
thought of as a type of M9-brane, with three special compact directions.

\ack
{I would like to thank
the Laboratoire de Physique Th\' eorique, Ecole Normale Sup\' erieure, where
early parts of this work were carried out,  for hospitality,
and Jos\' e Figueroa-O'Farrill  for useful discussions.}

\refout

\bye